\documentclass[prc,twocolumn,letterpaper,10pt,twoside,tightenlines,nofootinbib,showpacs]{revtex4}
\usepackage{amsmath,amsfonts,latexsym}
\usepackage{graphicx}
\usepackage[sort&compress]{natbib}
\begin{document}

\newcommand{\eg}{{\it e.g.}}
\newcommand{\etal}{{\it et. al.}}
\newcommand{\ie}{{\it i.e.}}
\newcommand{\be}{\begin{equation}}
\newcommand{\ee}{\end{equation}}
\newcommand{\bea}{\begin{eqnarray}}
\newcommand{\eea}{\end{eqnarray}}
\newcommand{\bef}{\begin{figure}}
\newcommand{\eef}{\end{figure}}
\newcommand{\bce}{\begin{center}}
\newcommand{\ece}{\end{center}}

\newcommand{\dd}{\text{d}}
\newcommand{\ii}{\text{i}}
\newcommand{\lsim}{\lesssim}
\newcommand{\gsim}{\gtrsim}

\title{Nonperturbative Heavy-Quark Diffusion in the Quark-Gluon Plasma}

\author{H. van Hees$^1$, M.~Mannarelli$^2$, V.Greco$^3$ and R.~Rapp$^1$}
\affiliation{$^1$Cyclotron Institute and Physics Department,
       Texas A\&M University, College Station, Texas 77843-3366, U.S.A.}
\affiliation{$^2$Instituto de Ciencias del Espacio (IEEC/CSIC), E-08193 Bellaterra
(Barcelona), Spain}
\affiliation{$^3$Dipartimento di Fisica e Astronomia, Via S. Sofia 64,
  I-95125 Catania, Italy}

\date{\today}

\begin{abstract}
  We evaluate heavy-quark (HQ) transport properties in a Quark-Gluon
  Plasma (QGP) within a Brueckner many-body scheme employing interaction 
  potentials extracted from thermal lattice QCD. The in-medium $T$-matrices
  for elastic charm- and bottom-quark scattering off light quarks in the QGP
  are dominated by
  attractive meson and diquark channels which support resonance states 
  up to temperatures of $\sim$1.5~$T_c$. The resulting drag coefficient 
  increases with decreasing temperature, contrary to expectations based 
  on perturbative QCD scattering.
  Employing relativistic Langevin simulations we compute HQ spectra 
  and elliptic flow in $\sqrt{s_{NN}}$=200~GeV Au-Au collisions. A good 
  agreement with electron decay data supports our nonperturbative 
  computation of HQ diffusion, indicative for a strongly coupled QGP.
\end{abstract}
\maketitle

Experiments at the Relativistic Heavy-Ion Collider (RHIC) have shown
that the matter produced in Au-Au collisions cannot be described by a
weakly interacting gas of quarks and gluons, but rather consists of a
strongly coupled Quark-Gluon Plasma (sQGP) with remarkably large opacity
and low viscosity. The latter is required by hydrodynamic descriptions
of the expanding fireball, implying rapid thermalization of the
medium~\cite{Shuryak:2003xe,Kolb:2003dz}. The understanding of these
properties in terms of the underlying interactions in the QGP, as
governed by Quantum Chromodynamics (QCD), is a key theoretical
objective.  A valuable probe of the sQGP are heavy quarks (charm and
bottom) which, due to their large mass, $m_Q$$\gg$$T_c$
($T_c$$\simeq$180~MeV: critical temperature~\cite{Karsch:2007vw}), are
believed to be sensitive to the processes that establish and maintain
thermalization of the medium, even at soft momentum scales. RHIC data
for single-electron ($e^\pm$) spectra associated with semileptonic
heavy-quark (HQ) decays in Au-Au collisions exhibit a surprisingly
strong suppression and elliptic
flow~\cite{Adler:2005xv,Abelev:2006db,Adare:2006nq}, indicating
substantial collective behavior of charm quarks in the expanding
fireball. Perturbative QCD (pQCD) calculations, based on radiative
energy loss, cannot explain these findings, even after inclusion of
elastic scattering~\cite{Armesto:2005mz,Wicks:2005gt}. Furthermore, it
has been argued that the convergence of the perturbative series for the
HQ diffusion constant is rather poor~\cite{CaronHuot:2007gq}, which
calls for nonperturbative approaches. Effective models with strong HQ
coupling in the 
QGP~\cite{vanHees:2004gq,Moore:2004tg,Zhu:2006er,Vitev:2007jj} 
lead to significantly reduced thermal relaxation times compared to pQCD 
elastic scattering~\cite{Svetitsky:1987gq}, resulting in better
agreement~\cite{Zhang:2005ni,vanHees:2005wb} with $e^\pm$
spectra~\cite{Adler:2005xv,Abelev:2006db,Adare:2006nq}.

In the present article, we perform a microscopic calculation of HQ
diffusion in the QGP employing a nonperturbative $T$-matrix
approach~\cite{Mannarelli:2005pz} with a driving kernel (potential)
estimated from finite-temperature lattice QCD computations. We include a
complete set of color channels for heavy-light quark interactions, as
well as $l$=0,1 partial waves together with HQ spin symmetry.  This, in
principle, provides an estimate of (elastic) transport coefficients
without tunable parameters, albeit significant uncertainties remain in
the definition of the potential. Within these uncertainties applications
to HQ observables at RHIC support our approach.

To evaluate in-medium properties of heavy quarks ($Q$=$c$,$b$) and
heavy-light quark correlations, we employ a Brueckner-type many-body
approach~\cite{Mannarelli:2005pz}. Our key assumptions are: (i) the main
features of the elastic heavy-light quark interaction can be
approximated by a static interaction potential, $V(r)$ (to leading order
in 1/$m_Q$, such an approach has been successfully applied for $D$ meson
spectra and decays in the vacuum~\cite{Avila:1994vi,Godfrey:1985xj}),
and (ii) $V(r)$ can be extracted from lQCD simulations of the
singlet-free energy $F_1(r,T)$~\cite{Kaczmarek:2003dp} for a static
$\bar QQ$ pair. As in previous
works~\cite{Shuryak:2004tx,Wong:2004zr,Mannarelli:2005pz,Cabrera:2006wh},
we identify the potential with the internal energy, $U_1$, which is
obtained by subtracting the entropy contribution from the free energy,
\begin{equation}
V_1(r,T)=U_1(r,T)-U_1(\infty,T) , \ \  
U_1\,=\,F_1-T\frac{{\rm d} F_1}{{\rm d} T} \ . 
\label{V1}
\end{equation}
A further subtraction is required to ensure the vanishing of the
potential at large distance and thus the convergence of the $T$-matrix
integral equation. In lQCD simulations the large distance limit of the
internal energy, $U_1(\infty,T$$>$$T_c)$, is a decreasing function of
the temperature. It is tempting to associate this quantity with a
selfenergy contribution to the HQ mass, $m_Q(T)$=$m_0 + U_1(\infty,T)/2$
($m_0$: ``bare'' mass). However, around $T_c$, $U_1(\infty,T)$ develops
a rather pronounced maximum structure rendering a mass interpretation
problematic. Furthermore, little is known about the momentum dependence
of this quantity. For simplicity, we assume constant values for
effective $c$- and $b$-quark masses of $m_c$=1.5~GeV and $m_b$=4.5~GeV
(the difference to the current mass is mainly attributed to perturbative 
contributions).

In addition to the color-singlet (meson) channel, we consider HQ
interactions in the color-octet $Q\bar q$, as well as in antitriplet
and sextet $Qq$ (diquark) channels. For the corresponding potentials we
adopt Casimir scaling according to leading-order (LO) perturbation
theory, $V_8 = - \frac{1}{8} V_1$, $V_{\bar 3} = \frac{1}{2} V_1$, 
$V_6 = - \frac{1}{4} V_1$, which is also supported by lQCD calculations 
of the finite-$T$ HQ free energy~\cite{Nakamura:2005hk,Doring:2007uh}.

The largest uncertainty in our calculations resides
in the definition and extraction of the potential. While the
identification with the internal energy (rather than the free energy)
may be considered as an upper limit, the variations between different
lQCD calculations and pertinent parametrizations to numerically evaluate
the entropy term in Eq.~(\ref{V1}), are appreciable. We have adopted 
3 different potentials, based on parametrizations of $F_1(r,T)$ in 
Refs.~\cite{Wong:2004zr}=[Wo],
\cite{Shuryak:2004tx}=[SZ] and \cite{Mannarelli:2005pz}=[MR] for
quenched~\cite{Kaczmarek:2003dp}, 2-flavor~\cite{Kaczmarek:2003ph} and
3-flavor lQCD~\cite{Petreczky:2004priv}, respectively. The
parametrizations [Wo] and [SZ] are similar to a recent
extraction~\cite{Cabrera:2006wh} from 3-flavor
lQCD~\cite{Petreczky:2004pz}. The [MR] potential is deeper than the
other two for $T$$\lsim$1.6~$T_c$ (and consequently gives larger
effects), but falls off faster above. The resulting transport
coefficients vary by $\sim$40\%. More details will be elaborated in an
extended paper~\cite{vanHees:2007xx}; in the following we restrict
ourselves to the [Wo] potential. The lQCD-based potentials are
implemented into a Brueckner many-body approach for heavy quarks,
defined by a system of coupled Bethe-Salpeter (BS) and Dyson equations:
\begin{alignat}{2}
T &= K + \int K G T \, ,
\label{BS0}  \\
\Sigma^Q &= \Sigma_g +\int\! T S^q
\label{Sigma0}, \quad
S^Q = S^Q_{0} + S^Q_{0} \Sigma^Q S^Q \, , 
\end{alignat}
with $T$: heavy-light quark $T$-matrix, $K$: interaction kernel, $G$:
2-particle propagator, $S^{Q,q}$ ($S_0^{Q,q}$): (free) single-particle
propagators for heavy and light quarks, $\Sigma^Q$: HQ selfenergy with
contributions from thermal gluons ($\Sigma_g$) and the $T$-matrix part
from interactions with thermal light-anti-/quarks.  Since we focus on a
QGP at zero chemical potential ($\mu_q$=0), all quantities are
quark-antiquark symmetric. To close the equations in the quark sector,
one needs the corresponding system of equations for the light sector,
which has been solved selfconsistently for $\Sigma_q$ and $T_{q\bar q}$
in Ref.~\cite{Mannarelli:2005pz}. Here, we augment these results by $qq$
diquark interactions and implement the quark selfenergies in simplified
form with constant real and imaginary parts as an effective quark mass,
$m_q$=0.25~GeV, and width, $\Gamma_q$=0.2~GeV~\cite{Mannarelli:2005pz};
variations in these parameters have little impact on the resulting quark
selfenergies. The effects from heavy quarks in the heat bath can be
safely neglected.

\begin{figure}[!t]
\includegraphics[width=0.9\linewidth]{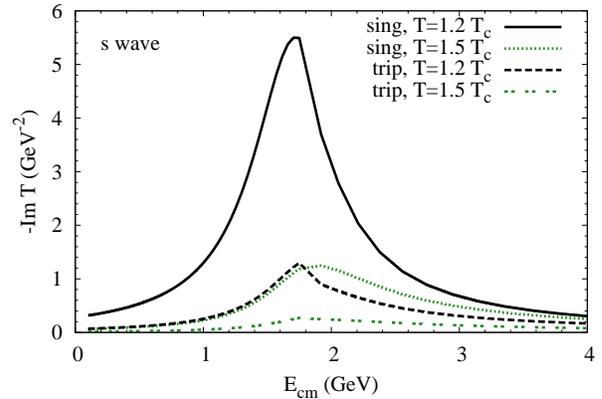}\hspace*{3mm}
\caption{Imaginary part of the in-medium $T$-matrix for $S-$wave $c\bar
  q$ and $cq$ scattering in color-singlet and antitriplet channels,
  respectively, at two different temperatures. The two-body threshold is
  at $E_{\rm thr}$=$m_c$+$m_q$=1.75~GeV.}
\label{fig_TQq}
\end{figure}
To implement the (static) lQCD-based potential into our Brueckner
approach we adopt the following approximations. First, we use a standard
reduction scheme~\cite{Thompson:1970wt} to convert the 4-dimensional BS
equation into a 3-D Lippmann-Schwinger (LS) equation (other
reductions~\cite{Blankenbecler:1965gx} lead to very similar
results~\cite{Mannarelli:2005pz}), thus neglecting virtual
particle-antiparticle loops but keeping relativistic kinematics of the
quark propagators. This, in turn, enables to identify the reduced kernel
$K$ with the potentials $V_{1,\bar 3,6,8}$ constructed above,
representing a ladder approximation to the $T$-matrix. As in
Refs.~\cite{Shuryak:2004tx,Mannarelli:2005pz}, we correct the static
potentials for a relativistic Breit (current-current) interaction.
Azimuthal symmetry and a partial-wave expansion lead to a 1-D LS
equation in each color ($a$) and angular-momentum ($l$) channel ($E$:
center-of-mass energy of the $Qq$ system),
\begin{equation}
\begin{split}
  T_{a,l}(E;q^\prime, q) = V_{a,l}( q^\prime, q) + \frac{2}{\pi} \int
  \dd k \, k^2 \, V_{a,l}( q^\prime, k) \quad \\
  \qquad\times G_{Qq}(E; k) T_{a,l}(E;k,q)
  [1-f(\omega^Q_k)-f(\omega^q_k)] \, .
\end{split}
\label{Tmatrix} 
\end{equation} 
$f(\omega)$ denote Fermi-Dirac distributions and
$\omega_k^i=(m_i^2+k^2)^{1/2}$ quasiparticle dispersion laws.  We
include both $S$- ($l$=0) and $P$-wave ($l$=1) channels.  The 2-particle
propagator in the Thompson scheme~\cite{Thompson:1970wt} reads
\begin{equation}
G(E; k) = (1/4)/[E-(\omega^q_k+{\rm i}\Sigma^q_I) -
(\omega^Q_k + {\rm i} \Sigma^Q_I)] \  . 
\label{GTh} 
\end{equation}
Results for the in-medium $S$-wave $T$-matrix are illustrated in 
Fig.~\ref{fig_TQq} for $c$-quark scattering. The attractive 
color-singlet and -antitriplet channels are the dominant 
contributions, supporting resonance structures up to temperatures 
of $\sim$1.7$T_c$ and $\sim$1.4$T_c$, respectively. Both the repulsive
color channels, as well as $P$-waves, lead to much
smaller $T$-matrices. However, due to larger degeneracies their 
contribution to the HQ selfenergies and transport coefficients is
not negligible.

\begin{figure}[!t]
\includegraphics[width=0.9\linewidth]{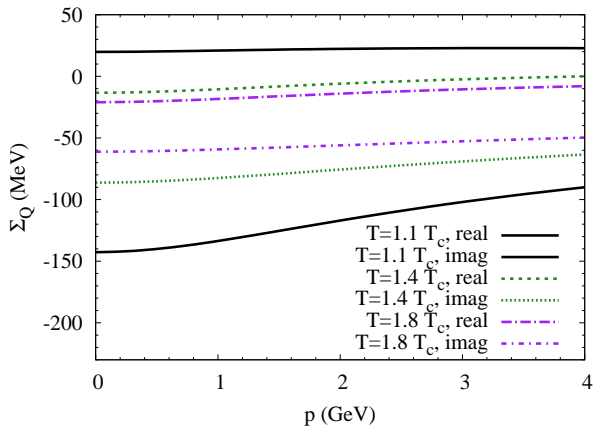}
\caption{Real (upper lines) and imaginary (lower lines) parts of the 
on-shell $c$-quark selfenergy versus 3-momentum at 
temperatures $T$=1.1, 1.4 and 1.8~$T_c$.}
\label{fig_self}
\end{figure}
Next we compute the HQ selfenergies and transport coefficients generated
by the nonperturbative heavy-light $T$-matrices. The HQ selfenergy
represented by the second term on the right-hand-side of
Eq.~(\ref{Sigma0}) is calculated within the imaginary-time formalism as
\begin{equation}
\begin{split}
\Sigma^Q_a(z_v;p) = & \frac{d_{SI} d_a}{6} \int \frac{\dd^3p'}{(2\pi)^3}
(-T) \\
& \times \sum\limits_{z_{\nu'}} T_a(z_\nu+z_{\nu'};\mathbf{p},\mathbf{p}')
D_{\bar q}(z_{\nu'},{\bf p}') 
\label{selfenergy}
\end{split}
\end{equation}
($z_\nu$=$\ii \pi$($2\nu$+1)$T$: fermionic Matsubara frequencies). As
implicit in our $T$-matrix (potential) we assume spin and light-flavor
degeneracy of the heavy-light interaction (in line with the free
$D$-meson spectrum~\cite{Abe:2003zm}), yielding $d_{SI}$=4(12)$N_f$ for
$S$($P$)-waves ($N_f$=2.5 to account for the smaller strange-quark
density). The resulting $c$-quark selfenergies (Fig.~\ref{fig_self})
translate into large in-medium widths of around
$\Gamma_c$=-2~Im$\Sigma_c$$\simeq$200~MeV (consistent with our input
parameters). The dominant meson and diquark contributions are about
equal, while the $P$-wave amounts to $\sim$40\%. The nonperturbative
real parts are small.

\begin{figure}[!t]
\includegraphics[width=0.9\linewidth]{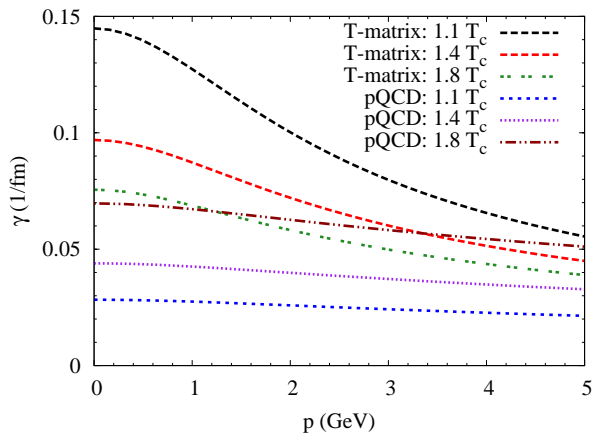}
\caption{Friction coefficients of charm quarks vs. 3-momentum  for
  different temperatures in a QGP, based on our nonperturbative
  $T$-matrix approach (upper curves at $p$=0) and on LO pQCD scattering 
  with $\alpha_s$=0.4 (lower curves).}
\label{fig_fpcoeff}
\end{figure}
We evaluate HQ diffusion in the QGP within a Fokker-Planck equation for
the distribution function, $f_Q$~\cite{Svetitsky:1987gq},
\begin{equation}
\frac{\partial f_Q}{\partial t} = \frac{\partial}{\partial p_i}
(p_i\gamma f_Q) + \frac{\partial^2}{\partial p_i \partial p_j}
(B_{ij} f_Q) \ ,
\end{equation}
with momentum and temperature dependent friction ($\gamma$) and
diffusion ($B_{ij}$) coefficients. They are calculated as in
Ref.~\cite{vanHees:2004gq} using the appropriate relation between the
invariant amplitude $\mathcal{M}$ and our in-medium $T$-matrix,
\begin{equation}
\begin{split}
\sum |\mathcal{M}|^2=\frac{64\pi}{s^2} (s-m_q^2+m_Q^2)^2(s-m_Q^2+m_q^2)^2 \\
\times N_f\sum_{a} d_a (|T_{a,l=0}(s)|^2 +3 |T_{a,l=1} (s)
\cos(\theta_{\text{cm}})|^2)\,.
\end{split}
\label{Msq}
\end{equation}
The non-perturbative thermal relaxation rates reach up to
$\gamma$$\simeq$1/(7fm/c) at low momenta close to $T_c$, a factor of
$\sim$4 larger than elastic pQCD scattering (but comparable to the
resonance model of Ref.~\cite{vanHees:2004gq}),
cf.~Fig.~\ref{fig_fpcoeff}. In contrast to other calculations available
thus far, the thermalization rate \textit{decreases} with temperature,
due to the dissolving resonances induced by the screening in the
lQCD-based potentials. The increase in quark density is overcompensated
by the loss of interaction strength. This has important consequences for
HQ observables at RHIC, as discussed below. The
$\cos^2(\theta_{\text{cm}})$ factor in $|\mathcal{M}|^2$ reduces the
$P$-wave contribution to $\gamma$ to $\sim$20\% of the $S$-wave.
Combining $T$-matrix and pQCD contributions, the spatial HQ-diffusion
constant at $p$=0 amounts to $D_{HQ}$=$T/(m_c\gamma)$$\simeq$5/2$\pi T$
at $T$=200~MeV, a factor of $\sim$4 smaller than in pQCD, thus
corroborating the notion of a strongly coupled QGP at temperatures up to
1.5-2~$T_c$.

\begin{figure}[!t]
\centerline{\includegraphics[width=0.9\linewidth]{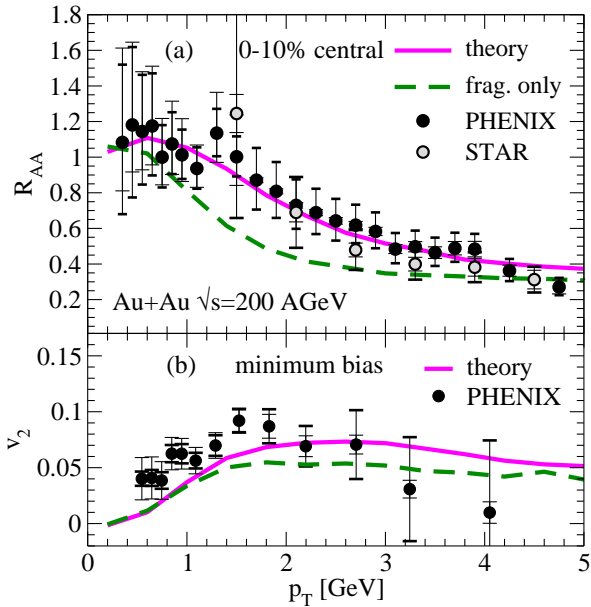}}
\caption{Our results for the nuclear modification factor (upper panel)
  and elliptic flow (lower panel) of single electrons with/without
  (solid/dashed lines) quark coalescence in Au-Au collisions compared to
  RHIC data~\cite{Abelev:2006db,Adare:2006nq}. The estimated theoretical
  uncertainty due to different extractions of the lQCD based heavy-light
  quark potentials is up to $\sim$$30\%$~\cite{vanHees:2007xx}.}
\label{fig_electron}
\end{figure}
The above calculated transport coefficients (from the $T$-matrix plus LO
pQCD scattering off gluons) are implemented into Langevin simulations of
$c$ and $b$ quarks in Au-Au collisions at RHIC using an isentropically
expanding QGP fireball~\cite{vanHees:2005wb}. The latter has been
constructed to resemble hydrodynamic models, with a bulk elliptic flow
of $\sim$5.5\% and initial temperature of $T_0$=340~MeV in semicentral
collisions (when using LO pQCD interactions, the resulting HQ spectra
are in good agreement with Langevin simulations in an explicit
hydrodynamic background~\cite{Moore:2004tg}). To compare to experimental
$e^\pm$ spectra, $c$ and $b$ quarks are hadronized at the end of the
QGP-hadron gas mixed phase within the coalescence model of
Ref.~\cite{Greco:2003vf} (supplemented by $\delta$-function
fragmentation and $D$- and $B$-meson decays). Note that the resonance
correlations in the $T$-matrix naturally merge into a quark-coalescence
description toward $T_c$~\cite{Ravagli:2007xx}.  Our calculations for
the $e^\pm$ nuclear modification factor, $R_{\rm AA}$ (defined as the
ratio of the spectrum in Au-Au collisions to the one in $p$-$p$ scaled
by the number of binary $N$-$N$ collisions), and elliptic flow
coefficient, $v_2$, show fair agreement with recent RHIC
data~\cite{Abelev:2006db,Adare:2006nq},
cf.~Fig.~\ref{fig_electron}. Inspection of the time evolution of the
$c$-quark distribution reveals that the suppression in the $p_T$ spectra
(i.e., $R_{AA}$) is mostly built up in the early stages of the QGP,
while most of the $v_2$ is developed at temperatures close to
$T_c$. This feature is amplified by the temperature dependence of the
transport coefficients in the $T$-matrix approach, and seems to be
favored by the $e^\pm$ data (coalescence with light quarks further
contributes to the increase in both $v_2$ and $R_{AA}$).


In summary, we have calculated HQ selfenergies and transport
coefficients within a $T$-matrix approach for heavy-light quark
interactions in the QGP using two-body potentials estimated from lattice
QCD. HQ scattering turns out to be dominated by ``prehadronic'' mesonic
and diquark channels which increase in strength when approaching $T_c$.
These correlations substantially accelerate thermal relaxation times
compared to pQCD and provide for a natural onset of the hadronization
process (in such a scenario, nonperturbative HQ interactions with gluons
are less relevant).  When implemented into Langevin simulations at RHIC,
reasonable agreement with the suppression and elliptic flow of $e^\pm$
spectra from HQ decays emerges. This is rather remarkable in view of the
largely parameter-free calculation of the transport coefficients. Future
work should aim at scrutinizing the uncertainties inherent in the
potential approach at finite temperature and in the extraction of the
potential from lattice QCD. Further insights could be obtained from
direct lQCD computations of heavy-light quark correlation functions in
the QGP. In addition, elastic HQ interactions, which parametrically
dominate at low $p_T$, should be supplemented by radiative energy
loss~\cite{Vitev:2007jj} which takes over at high $p_T$.  Our present
analysis suggests that a small HQ diffusion coefficient arises from a
nonperturbative interaction strength in a strongly coupled QGP.

This work has been supported by a U.S. National Science
Foundation CAREER Award under grant PHY-0449489 (HvH, RR), 
and by the Ministerio de Educaci\'on y Ciencia 
under grant AYA 2005-08013-C03-02 (MM).

\end{document}